\documentclass[twocolumn]{aastex62}
\usepackage{amsmath,color,textcomp,url,graphicx,subfigure}
\usepackage{morefloats}

\newcommand{\nsample}{708}

\newcommand{\newcms}{219}
\newcommand{\news}{254}
\newcommand{\confirmedcms}{66}
\newcommand{\reclassedconfirmed}{9}

\newcommand{\missedbfs}{53}
\newcommand{\bfide}{32}
\newcommand{\bfidetable}{38}

\newcommand{\confirmedcmstxt}{Sixty six}

\newcommand{\confirmedplusreclassedcmstxt}{Seventy five}

\newcommand{\bfidetabletxt}{Thirty eight}

\newcommand{\nasso}{27}
\newcommand{\gaia}{TGAS}
\newcommand{\bsigma}{\cite{2018ApJ...856...23G}}
\newcommand{\bsigmap}{\citep{2018ApJ...856...23G}}
\newcommand{\bsigmalp}{\citealp{2018ApJ...856...23G}}
\newcommand{\kms}{\hbox{km\,s$^{-1}$}}

\newcommand{\msol}{$M_{\odot}$}
\newcommand{\masyr}{$\mathrm{mas}\,\mathrm{yr}^{-1}$}

\usepackage{color}
\definecolor{myred}{RGB}{200,0,0}

\newenvironment{changemargin}[1]{%
\begin{list}{}{%
\setlength{\textheight}{#1}%
}%
\item[]}{\end{list}}

\submitjournal{ApJ}

\shorttitle{New Young Association Members in \emph{Gaia}--Tycho}
\shortauthors{Gagn\'e et al.}

\begin{document}

\title{BANYAN. XII. NEW MEMBERS OF NEARBY YOUNG ASSOCIATIONS FROM \emph{GAIA}--TYCHO DATA}

\author[0000-0002-2592-9612]{Jonathan Gagn\'e}
\affiliation{Carnegie Institution of Washington DTM, 5241 Broad Branch Road NW, Washington, DC~20015, USA}
\affiliation{NASA Sagan Fellow}
\email{jgagne@carnegiescience.edu}
\author[0000-0003-2005-5626]{Olivier Roy-Loubier}
\affil{Institute for Research on Exoplanets, Universit\'e de Montr\'eal, D\'epartement de Physique, C.P.~6128 Succ. Centre-ville, Montr\'eal, QC H3C~3J7, Canada}
\author[0000-0001-6251-0573]{Jacqueline K. Faherty}
\affiliation{Department of Astrophysics, American Museum of Natural History, Central Park West at 79th St., New York, NY 10024, USA}
\author[0000-0001-5485-4675]{Ren\'e Doyon}
\affil{Institute for Research on Exoplanets, Universit\'e de Montr\'eal, D\'epartement de Physique, C.P.~6128 Succ. Centre-ville, Montr\'eal, QC H3C~3J7, Canada}
\author[0000-0002-8786-8499]{Lison Malo}
\affil{Observatoire du Mont-M\'egantic and Institute for Research on Exoplanets, Universit\'e de Montr\'eal, D\'epartement de Physique, C.P.~6128 Succ. Centre-ville, Montr\'eal, QC H3C~3J7, Canada}

\begin{abstract}

We present a search for stellar members of young associations within 150\,pc of the Sun based on \gaia\ and an updated version of the BANYAN~$\Sigma$ software to determine Bayesian membership probabilities that includes \emph{Gaia}--2MASS color-magnitude diagrams. We identify \bfide\ new F0--M3-type bona fide members of the 10--200\,Myr-old Sco-Cen, Carina, Tucana-Horologium, Columba and Octans associations and the AB~Doradus, $\beta$\,Pictoris and Carina-Near moving groups. These new bona fide members have measurements of their full kinematics and literature data consistent with a young age. We also confirm the membership of \confirmedcms\ previously known candidate members using their \emph{Gaia}--Tycho trigonometric distances or new literature radial velocities, and identify \newcms\ additional new candidate members, most of which do not yet have a radial velocity measurement. This work is the first step towards a completeness-corrected survey of young association members based on \emph{Gaia}--DR2 in the near future.

\end{abstract}


\section{INTRODUCTION}\label{sec:intro}

Young associations and moving groups in the Solar neighborhood are valuable laboratories to study the properties of age-calibrated stars in detail \citep{2004ARAA..42..685Z,2008hsf2.book..757T}. Their proximity makes it possible to gather high-quality data more easily, perform high-angular resolution imaging of exoplanets and bring this characterization to the lowest-mass and faintest substellar objects with a calibrated age (e.g., see \citealp{2008Sci...322.1348M,2013ApJ...777L..20L,2013AA...553L...5D,2014ApJ...787....5N,2015Sci...350...64M,2015ApJS..219...33G,2016ApJS..225...10F,2017ApJ...841L...1G}). However, the main difficulty in studying such associations is also a consequence of their proximity: their members are distributed over large areas of the sky (e.g., see Figure~1 of \bsigmalp), making it difficult to identify members without measuring their full 6-dimensional kinematics, consisting of the $XYZ$ Galactic coordinates and $UVW$ space velocities. These require measuring the trigonometric parallax and heliocentric radial velocities of stars, which is challenging to perform on all stars in a large area of the sky.

Until recently, the discovery and kinematic characterization of most young associations and moving groups in the Solar neighborhood relied on parallaxes obtained by the Hipparcos mission \citep{1997AA...323L..49P}, allowing us to discover only the most massive and brightest members of these young associations, in general within $\sim$\,100\,pc. The stars bright enough to have a Hipparcos parallax measurement only represent $\sim$\,10\% of the total stellar population within 30\,pc, and 2\% within 100\,pc \citep{1997AA...323L..49P,2013ApJS..208....9P}, which is a consequence of the initial mass function peaking at mass of $\sim$\,0.25\,\msol\ \citep{2010AJ....139.2679B} corresponding to the M spectral class. Recent studies focused on various statistical methods to obtain the full kinematics of only the most likely members of these associations, based on their sky position, proper motion and photometry\added{, as well as radial velocities and parallaxes when available} (e.g., see \citealp{2005ApJ...634.1385M,2011ApJ...727...62R,2013ApJ...762...88M,2014ApJ...788...81M,2014AJ....147..146K,2014ApJ...783..121G,2015MNRAS.447.1267M,2015ApJ...798...73G,2017AJ....153...95R,2017AJ....154...69S}), and have started to uncover and confirm the kinematics of a fraction of their low-mass stars and substellar objects. The \emph{Gaia} mission \citep{2016AA...595A...1G} has already started to benefit this field of research with the Data Release 1 (\emph{Gaia}--DR1; \citealp{2016AA...595A...2G}), which provided 2 million parallaxes for the Tycho catalog\added{ \citep{2000AA...355L..27H}\footnote{Tycho has a \emph{Gaia} limiting magnitude of $G \approx 11.6$ (50\% recovery), and recovers less than 1\% of objects with $G \geq 12.7$.}}. This will be even more true of \emph{Gaia}--DR2, which will provide a billion parallaxes that will allow to complete the population of all young associations within 150\,pc down to $\sim$\,0.12\,\msol\ \citep{2017MNRAS.469..401S}. This completion of the stellar population of young associations in the Solar neighborhood will have many direct applications, such as comparing their initial mass functions (e.g., see \citealp{2005ASSL..327...41C,2007ApJS..173..104L,2011AJ....141...98B,2012EAS....57...45J,2017ApJS..228...18G}), providing strategic targets for the direct imaging of exoplanets, understanding the stellar formation history of the Solar neighborhood, and will provide important benchmark populations to characterize the fundamental properties and chemical abundances of coeval stars (e.g., see \citealp{2000ApJ...533..944K,2006ApJ...636..432S}).

The latest membership classification tool, BANYAN~$\Sigma$\added{ \bsigmap}, benefitted from \emph{Gaia}--DR1 data to refine its kinematic models of the \nasso\ well-characterized young associations within 150\,pc of the Sun. The ages of these associations are in the range $\sim$\,1--850\,Myr, and their general characteristics are listed in Table~\ref{tab:nyas}, which is a summarized version of Table~1 from \bsigma. The BANYAN~$\Sigma$ tool uses Bayesian inference to determine the membership probability that a star belongs to any of these \nasso\ associations, based on its sky position, proper motion, and optionally its radial velocity and distance. When radial velocity and/or distance are not known, these observables are marginalized and a membership probability is still calculated. BANYAN~$\Sigma$ includes more associations and generates less contaminants at a fixed recovery rate compared to all previous tools available in the literature\added{ (e.g., \citealp{2014ApJ...788...81M,2014ApJ...783..121G,2017AJ....153...95R})}. Furthermore, its analytical solving of marginalization integrals makes it less computationally intensive, and amenable to analyze much more easily large data sets such as \emph{Gaia}--DR1 as well as the upcoming data releases.

\startlongtable
\tabletypesize{\scriptsize}
\tablewidth{0.485\textwidth}
\setlength{\tabcolsep}{3pt}
\begin{deluxetable}{lccccccc}
\tablecolumns{7}
\tablecaption{Young associations included in this study.\label{tab:nyas}}
\tablehead{\colhead{Group} & \colhead{$\left<\varpi\right>$\tablenotemark{a}} & \colhead{$\left<\nu\right>$\tablenotemark{b}} & \colhead{$S_{\rm spa}$\tablenotemark{c}} & \colhead{$S_{\rm kin}$\tablenotemark{d}} & \colhead{Age} & \colhead{Ref.}\\
\colhead{Name} & \colhead{(pc)} & \colhead{(\kms)} & \colhead{(pc)} & \colhead{(\kms)} & \colhead{(Myr)} & \colhead{} }
\startdata
118TAU & $100 \pm 10$ & $14 \pm 2$ & 3.4 & 2.1 & $\sim$\,10 & 1\\
ABDMG & $30_{-10}^{+20}$ & $10_{-20}^{+10}$ & 19.0 & 1.4 & $149_{-19}^{+51}$ & 2\\
$\beta$PMG & $30_{-10}^{+20}$ & $10 \pm 10$ & 14.8 & 1.4 & $24 \pm 3$ & 2\\
CAR & $60 \pm 20$ & $20 \pm 2$ & 11.8 & 0.8 & $45_{-7}^{+11}$ & 2\\
CARN & $30 \pm 20$ & $15_{-10}^{+7}$ & 14.0 & 2.1 & $\sim$\,200 & 3\\
CBER & $85_{-5}^{+4}$ & $-0.1 \pm 0.8$ & 3.6 & 0.5 & $562_{-84}^{+98}$ & 4\\
COL & $50 \pm 20$ & $21_{-8}^{+3}$ & 15.8 & 0.9 & $42_{-4}^{+6}$ & 2\\
CRA & $139 \pm 4$ & $-1 \pm 1$ & 1.5 & 1.7 & 4--5 & 5\\
EPSC & $102 \pm 4$ & $14 \pm 3$ & 2.8 & 1.8 & $3.7_{-1.4}^{+4.6}$ & 6\\
ETAC & $95 \pm 1$ & $20 \pm 3$ & 0.6 & 2.0 & $11 \pm 3$ & 2\\
HYA & $42 \pm 7$ & $39_{-4}^{+3}$ & 4.5 & 1.2 & $750 \pm 100$ & 7\\
IC2391 & $149 \pm 6$ & $15 \pm 3$ & 2.2 & 1.4 & $50 \pm 5$ & 8\\
IC2602 & $146 \pm 5$ & $17 \pm 3$ & 1.8 & 1.1 & $46_{-5}^{+6}$ & 9\\
LCC & $110 \pm 10$ & $14 \pm 5$ & 11.6 & 2.2 & $15 \pm 3$ & 10\\
OCT & $130_{-20}^{+30}$ & $8_{-9}^{+8}$ & 22.4 & 1.3 & $35 \pm 5$ & 11\\
PL8 & $130 \pm 10$ & $22 \pm 2$ & 5.0 & 1.1 & $\sim$\,60 & 12\\
PLE & $134 \pm 9$ & $6 \pm 2$ & 4.1 & 1.4 & $112 \pm 5$ & 13\\
ROPH & $131 \pm 1$ & $-6.3 \pm 0.2$ & 0.7 & 1.6 & $<$\,2 & 14\\
TAU & $120 \pm 10$ & $16 \pm 3$ & 10.7 & 3.6 & 1--2 & 15\\
THA & $46_{-6}^{+8}$ & $9_{-6}^{+5}$ & 9.1 & 0.8 & $45 \pm 4$ & 2\\
THOR & $96 \pm 2$ & $19 \pm 3$ & 3.9 & 2.1 & $22_{-3}^{+4}$ & 2\\
TWA & $60 \pm 10$ & $10 \pm 3$ & 6.6 & 1.5 & $10 \pm 3$ & 2\\
UCL & $130 \pm 20$ & $5 \pm 5$ & 17.4 & 2.5 & $16 \pm 2$ & 10\\
UCRA & $147 \pm 7$ & $-1 \pm 3$ & 4.5 & 1.8 & $\sim$\,10 & 16\\
UMA & $25.4_{-0.7}^{+0.8}$ & $-12 \pm 3$ & 1.2 & 1.3 & $414 \pm 23$ & 17\\
USCO & $130 \pm 20$ & $-5 \pm 4$ & 9.9 & 2.8 & $10 \pm 3$ & 10\\
XFOR & $100 \pm 6$ & $19 \pm 2$ & 2.6 & 1.3 & $\sim$\,500 & 18\\
\enddata
\tablenotetext{a}{Peak of distance distribution and $\pm$1$\sigma$ range.}
\tablenotetext{b}{Peak of radial velocity distribution and $\pm$1$\sigma$ range.}
\tablenotetext{c}{Characteristic spatial scale in $XYZ$ space.}
\tablenotetext{d}{Characteristic kinematic scale in $UVW$ space.}
\tablewidth{0.4\textwidth}
\tablecomments{The full names of young associations are: 118~Tau (118TAU), AB~Doradus (ABDMG), $\beta$~Pictoris ($\beta$PMG), Carina (CAR), Carina-Near (CARN), Coma Berenices (CBER), Columba (COL), Corona~Australis (CRA), $\epsilon$~Chamaeleontis (EPSC), $\eta$~Chamaeleontis (ETAC), the Hyades cluster (HYA), Lower Centaurus Crux (LCC), Octans (OCT), Platais~8 (PL8), the Pleiades cluster (PLE), $\rho$~Ophiuci (ROPH), the Tucana-Horologium association (THA), 32~Orionis (THOR), TW~Hya (TWA), Upper Centaurus Lupus (UCL), Upper~CrA (UCRA), the core of the Ursa~Major cluster (UMA), Upper~Scorpius (USCO), Taurus (TAU) and $\chi^1$~For (XFOR).}
\tablerefs{(1)~\citealt{mamajek118tau}; (2)~\citealt{2015MNRAS.454..593B}; (3)~\citealt{2006ApJ...649L.115Z}; (4)~\citealt{2014AA...566A.132S}; (5)~\citealt{2012MNRAS.420..986G}; (6)~\citealt{2013MNRAS.435.1325M}; (7)~\citealt{2015ApJ...807...24B}; (8)~\citealt{2004ApJ...614..386B}; (9)~\citealt{2010MNRAS.409.1002D}; (10)~\citealt{2016MNRAS.461..794P}; (11)~\citealt{2015MNRAS.447.1267M}; (12)~\citealt{1998AJ....116.2423P}; (13)~\citealt{2015ApJ...813..108D}; (14)~\citealt{2008hsf2.book..351W}; (15)~\citealt{1995ApJS..101..117K}; (16)~\citealt{2018arXiv180109051G}; (17)~\citealt{2015AAS...22511203J}; (18)~\citealt{2010AA...514A..81P}.}
\end{deluxetable}

In this paper, we apply the BANYAN~$\Sigma$ tool to the Tycho catalog stars that benefit from parallaxes in the \replaced{\emph{Gaia}--DR1 (hereafter referered to as \gaia)}{Tycho--\emph{Gaia} Astrometric Solution (\gaia\ hereafter)} to identify \bfide\ new bona fide members and \newcms\ new candidate members of the \nasso\ nearest young associations. In Section~\ref{sec:cmd}, we build \emph{Gaia}--2MASS $M_G$ versus $G - J$ color-magnitude sequences for field stars and young associations of different age categories that will complement the kinematic analysis of BANYAN~$\Sigma$ in our determination of Bayesian membership probabilities. Our method for selecting new candidate members is described in Section~\ref{sec:candidates}, and we investigate their signs of youth such as UV, X-ray emission and model isochrones in Section~\ref{sec:youth}. The conclusion of this work is presented in Section~\ref{sec:conclusion}.

\section{\emph{GAIA}--2MASS COLOR-MAGNITUDE SEQUENCES OF YOUNG AND FIELD STARS}\label{sec:cmd}

\begin{figure}
	\centering
	\includegraphics[width=0.485\textwidth]{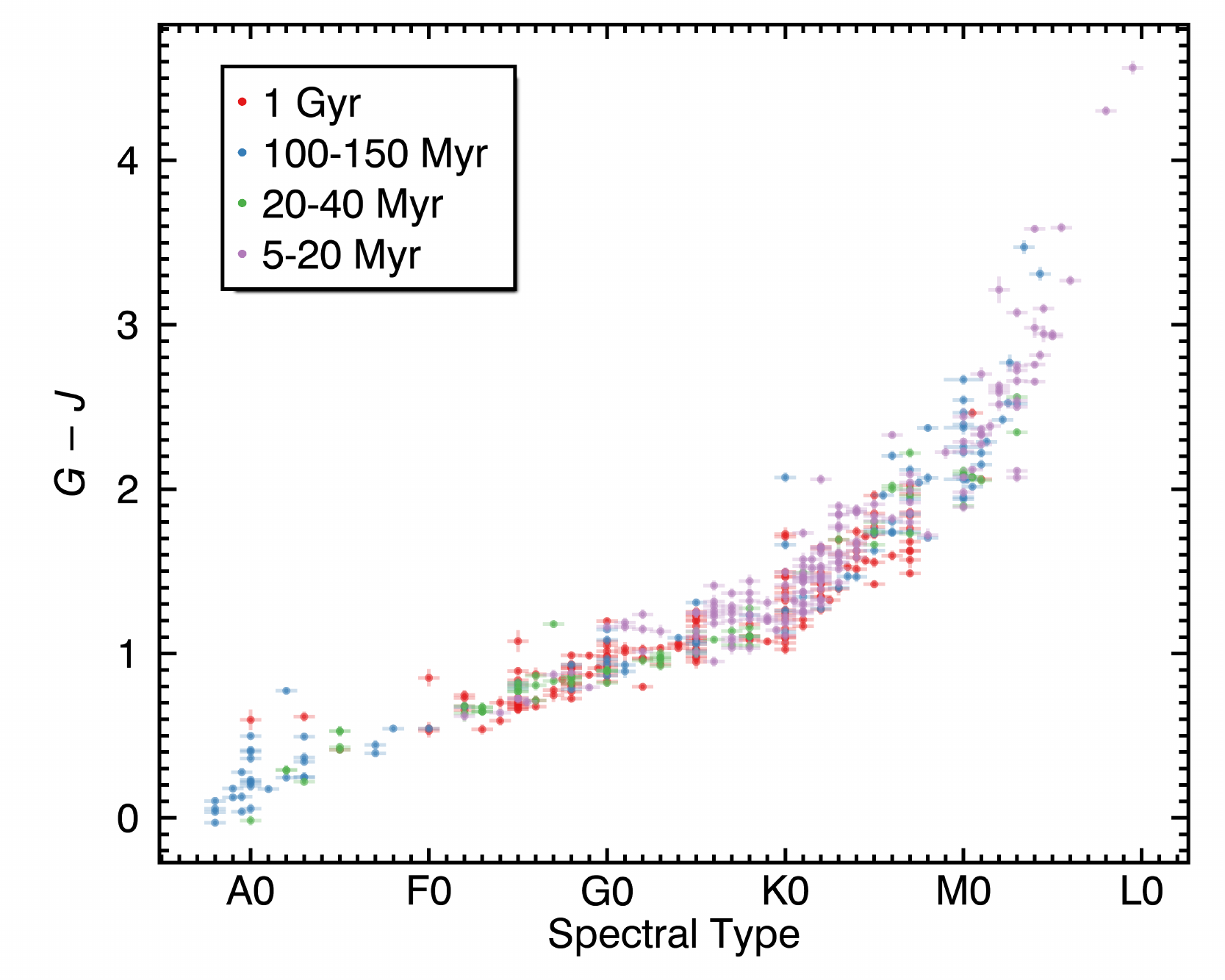}
	\caption{\emph{Gaia}--2MASS $G - J$  color as a function of spectral type for the bona fide members of young associations compiled by \bsigma, grouped by age. This color allows to separate spectral types well in the range A0--L0 and uses magnitudes that are readily available for a large number of stars. See Section~\ref{sec:cmd} for more detail.}
	\label{fig:spt_gj}
\end{figure}

In this section, color-magnitude sequences are built for field stars and members of young moving groups at different ages. In order for the sequence to be useful for the most stars and across a large range of spectral types regardless of age, it is preferrable to use a color combination that spans a wide wavelength window. The \emph{Gaia} $G$-band\added{\footnote{The \emph{Gaia} $G$ band has an effective central wavelength of $\sim$\,5900\,\AA\ and an effective width of $\sim$\,4200\,\AA\ as reported on \url{http://svo2.cab.inta-csic.es/svo/theory/fps/index.php?id=GAIA/GAIA0.G}, essentially encompassing the SDSS $griz$ bands.}} and 2MASS \citep{2006AJ....131.1163S} $J$-band magnitudes respect this criterion and are both readily available for a large number of stars. In Figure~\ref{fig:spt_gj}, we show the $G - J$ color as a function of spectral type for all bona fide members of young associations compiled by \bsigma, for which \gaia\ and 2MASS cross-matches were already provided. The members are grouped by ages to demonstrate that the spread in $G - J$ color is not significant in any age category. This figure demonstrates how the $G - J$ color provides a good proxy for spectral type in the A0--L0 range.

\begin{figure}
	\centering
	\includegraphics[width=0.485\textwidth]{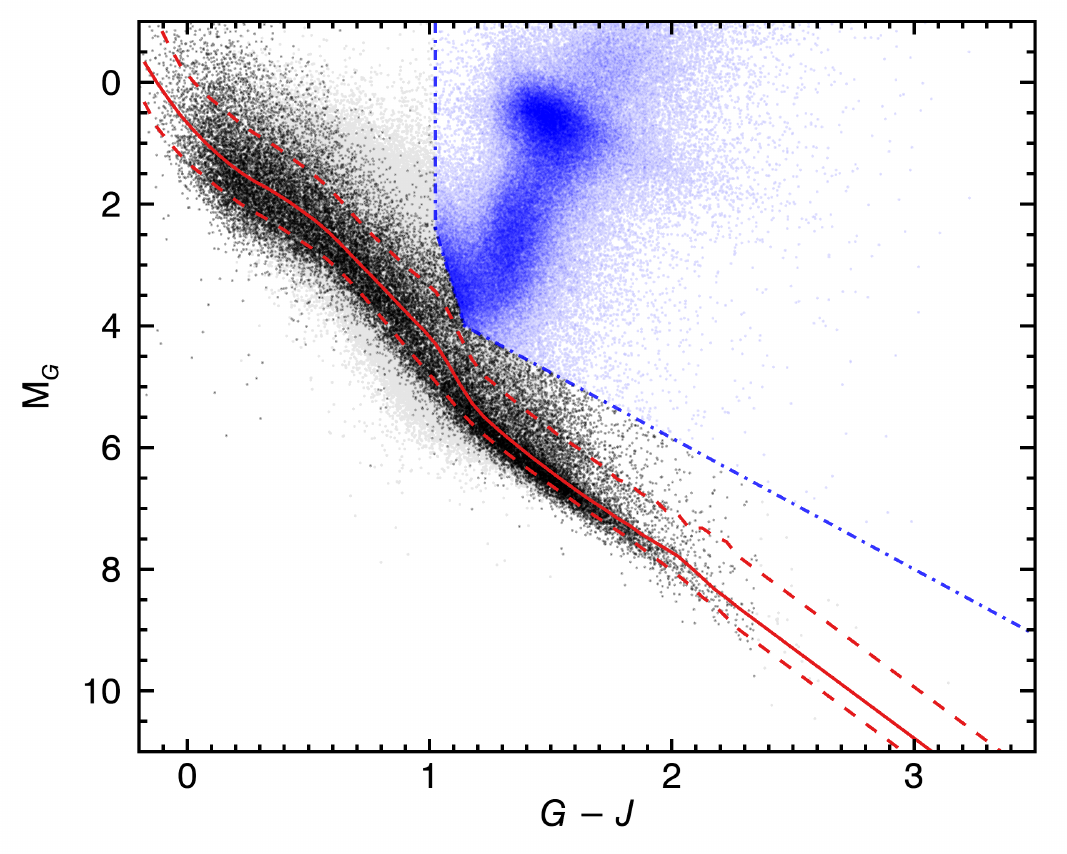}
	\caption{\emph{Gaia}--2MASS Color-magnitude sequence of field stars (red line) built from \gaia\ data (black dots). Giant stars rejected from the criterion described in Equation~\eqref{eqn:giants} are displayed as blue dots\added{ (delimited by the blue dash-dotted region)}. The red dashed lines represent the $\pm$\,1$\sigma$ range of the vertical distribution of \gaia\ data around the field sequence. The $G - J$ colors of most stars in \gaia\ are concentrated in the range 0.6--1.2\,mag, therefore here we displayed a fixed number of field stars (1000) per 0.05\,mag color bin to facilitate viewing. See Section~\ref{sec:cmd} for more detail.}
	\label{fig:cmd_field}
\end{figure}

The full data set of \gaia\ was used to build a color-magnitude sequence in absolute $M_G$ band as a function of $G - J$ color. The \gaia\ catalog entries were cross-matched with the nearest 2MASS entries using a search radius of 4\arcsec\ to achieve this. In order to build a monotonic sequence that can be described as a function in this color-magnitude space, giant stars were filtered out from the \gaia\ data. A simple color-magnitude region cut was used to achieve this (displayed in Figure~\ref{fig:cmd_field}); all stars redder or brighter than a color-magnitude region described by the three straight lines connecting these four ($G-J$,$M_G$) coordinates were rejected:
\begin{align}
	p_1 &= \left(1.02, -\infty\right),\notag\\
	p_2 &= \left(1.02, 2.40\right),\label{eqn:giants}\\
	p_3 &= \left(1.14, 4.00\right),\notag\\
	p_4 &= \left(3.00, 8.00\right).\notag
\end{align}

This rejection of giants from the field color-magnitude sequence means that a search for young stars that is not using parallax information may be susceptible to giant stars contaminating a sample of young stellar candidate members. This will however be mitigated by the kinematic analysis of BANYAN~$\Sigma$ that will reject low-proper motion objects unless they happen to match those of a young moving group by chance. In the current analysis, all entries of \gaia\ that fall in the region described in Equation~\eqref{eqn:giants} are ignored.

All stars in \gaia\ were first split in $G - J$ color bins of 0.05\,mag in the range -0.2--3.5\,mag to build the field sequence. A first sequence was built by measuring the median absolute $M_G$-band magnitude in each $G - J$ color bin. A cumulative distribution function of the absolute $M_G$-band magnitudes of all stars within each bin was then built, and used to determine the positive and negative error bars in absolute $M_G$-band magnitude that would each encompass 34\% of the population on both sides of the median value, such that both error bars contain 68\% of all data and correspond to a $\pm$\,1$\sigma$ range. All three resulting median and $\pm$\,1$\sigma$ color-magnitude sequences were then smoothed with a 2-cells wide running average over all color bins. The resulting color-magnitude sequence of field stars is presented in Figure~\ref{fig:cmd_field}.

\begin{figure}
	\centering
	\includegraphics[width=0.485\textwidth]{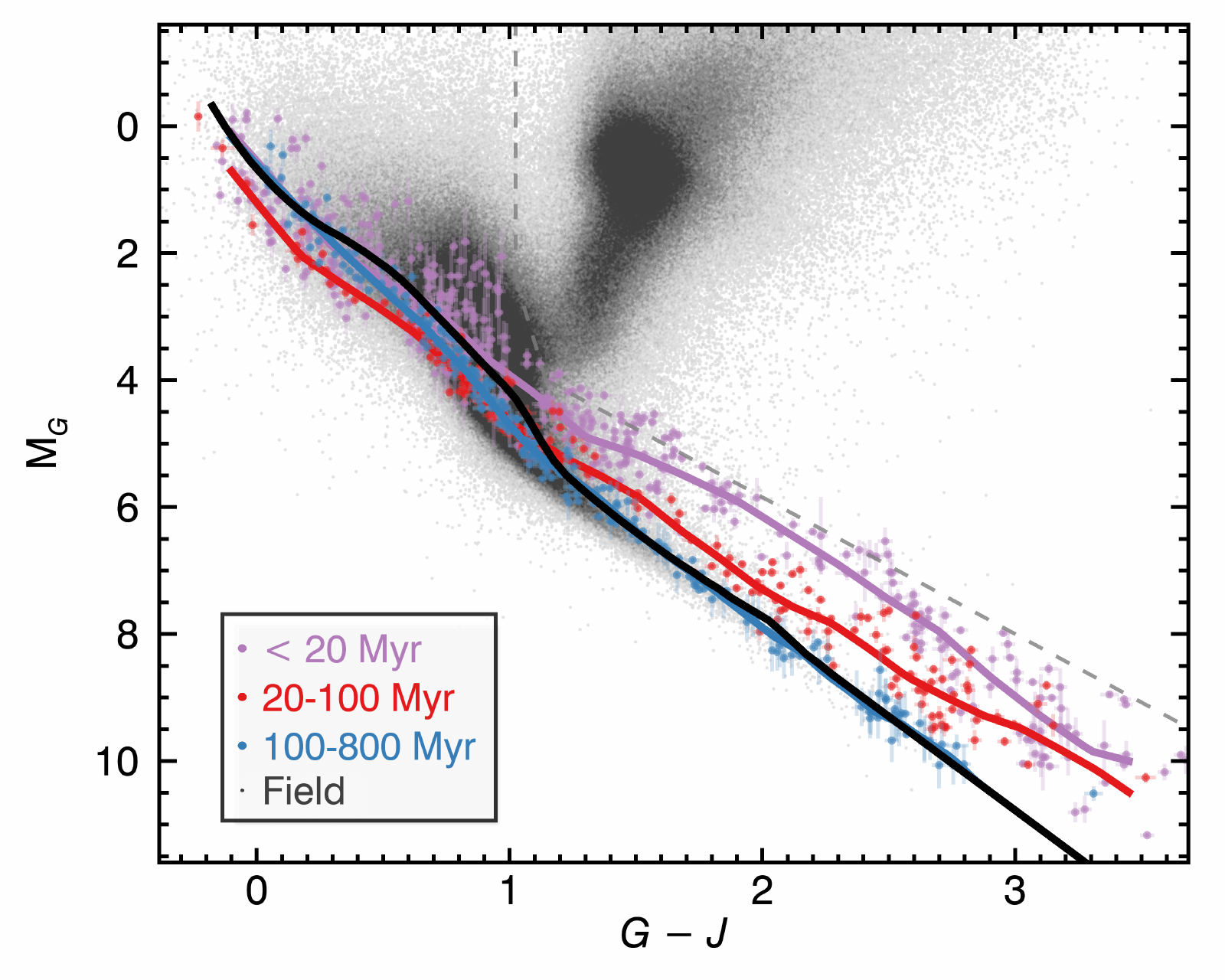}
	\caption{Color-magnitude sequences (thick lines) for the bona fide members of young associations compiled in \bsigma\ (colored circles). The \nasso\ young associations were grouped in three age categories which were found to follow distinct sequences in this particular color-magnitude diagram. \gaia\ entries are displayed as black dots.\added{ The gray dashed lines delimitate the giants exclusion criterion described in Equation~\eqref{eqn:giants}.} See Section~\ref{sec:cmd} for more detail.}
	\label{fig:cmd_young}
\end{figure}

Similar color-magnitude sequences were then built for the bona fide members of young associations compiled by \bsigma\added{\ (although the giants rejection criterion was not applied, as we have detailed literature information on these stars confirming that none of them are giants)}. All members were first assigned to one of three age categories, which were found to assemble the stars that follow similar color-magnitude sequences: (1) younger than 20\,Myr (i.e., members of 118TAU, CRA, EPSC, ETAC, LCC, ROPH, TAU, TWA, UCL, UCRA and USCO); (2) 20--100\,Myr (i.e., members of $\beta$PMG, CAR, COL, IC2602, IC2391, OCT, PL8, THA and THOR); (3) 100--800\,Myr (i.e., members of ABDMG, CARN, CBER, HYA, PLE, UMA and XFOR). \added{Known unresolved binaries were not included here, which will make the current search less sensitive in discovering binary systems in young associations\footnote{A version of BANYAN~$\Sigma$ with binary hypotheses will be released in a future publication.}.} The resulting color-magnitude diagrams are displayed in Figure~\ref{fig:cmd_young}. The 100--800\,Myr sequence is fainter than the field sequence at colors bluer than $G - J$ $\approx$ 1.0--1.2, and merges with it at redder colors. This can be attributed to field stars more massive than 0.96--1.00\,\msol\ (corresponding to $G - J$\,$\sim$\,1--1.2; \citealt{2013ApJS..208....9P}\footnote{See \url{http://www.pas.rochester.edu/~emamajek/EEM_dwarf_UBVIJHK_colors_Teff.txt}}) that start to depart from the main sequence onto the giant branch after $\sim$\,8--9\,Gyr \citep{2016ApJ...823..102C}. At younger ages of 20--100\,Myr, stars redder than $G - J$ $\approx$ 1.0--1.2 are brighter than the field sequence because their radii are still inflated from their young age (e.g., see \citealt{2014prpl.conf..219S}). This effect is more dramatic for stars younger than 20\,Myr, to the point where their sequence merges with that of the field at colors bluer than $G - J$ $\approx$ 1.0--1.2. This illustrates how stars coming into and departing from the main sequence are hard to distinguish using isochrones alone.

\section{IDENTIFICATION OF CANDIDATE MEMBERS}\label{sec:candidates}

The BANYAN~$\Sigma$ software uses sky position, proper motion and optionally radial velocity and distance to assign a Bayesian probability that a star is a member of a known young association within 150\,pc of the Sun. \gaia\ includes parallax measurements, sky position, proper motion and $G$-band magnitudes for 2\,057\,050 stars, but does not contain radial velocity measurements. We used the same method described in Section~\ref{sec:cmd} to cross-match \gaia\ entries with 2MASS in order to obtain the $J$-band magnitudes of all stars. Giant stars were filtered out from this sample with the criterion described in Section~\ref{sec:cmd}\added{ (only for stars with $M_G < 3$ to avoid rejecting low-mass stars younger than $\sim$\,20\,Myr.)}, resulting in a sample of 1\,338\,580 stars. All stars with a trigonometric distance measurement at a statistical significance of less than 2$\sigma$\added{ (i.e., measurement over error less than 2)} or with a missing 2MASS $J$-band measurement were rejected, further refining the sample to 1\,190\,699 stars.

All color-magnitude sequences described in Section~\ref{sec:cmd} were used to assign a Bayesian probability in each young association directly from a comparison of their $G - J$ color and absolute $G$-band magnitude, assuming that the vertical spread around each sequence is Gaussian. These probabilities are calculated by BANYAN~$\Sigma$ through the \texttt{constraint\char`_dist\char`_per\char`_hyp} and \texttt{constraint\char`_edist\char`_per\char`_hyp} keywords, and are subsequently multiplied to the Bayesian prior probabilities in determining membership probabilities (see \citealp{2018arXiv180109051G,zenodobanyansigmaidl,zenodobanyansigmapython} for more detail). The same color-magnitude sequences could be used to constrain the acceptable distances of any star without a parallax measurement, by using the same keyword in the BANYAN~$\Sigma$ software; it then automatically determines which method to adopt depending on whether trigonometric distances are available or not.

\begin{figure}
	\centering
	\includegraphics[width=0.465\textwidth]{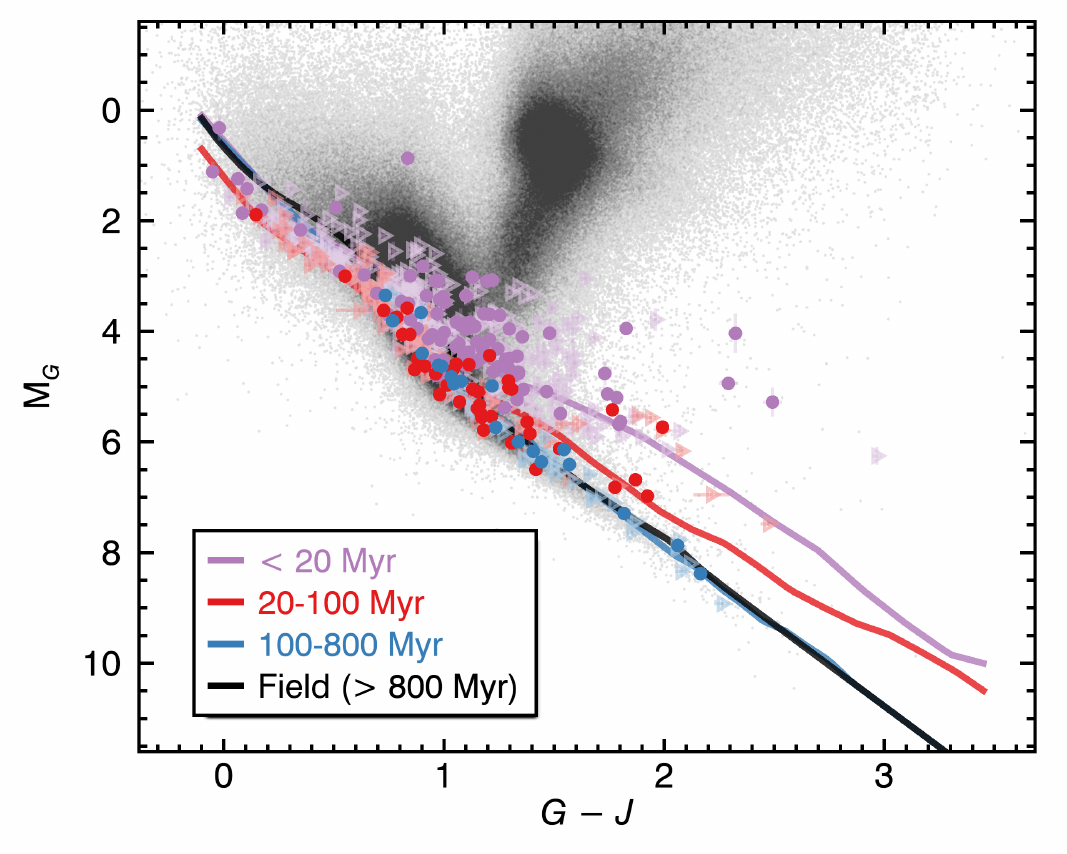}
	\caption{Color-magnitude diagram of all new candidate members identified here (open right triangles are objects without radial velocities; filled circles are objects with radial velocities) compared with the full \gaia\ catalog (black dots), and the field and young sequences (thick lines). The color scheme is identical to Figure~\ref{fig:cmd_young}. Outliers to their respective sequences are stars with disks and are discussed in Section~\ref{sec:candidates}.}
	\label{fig:all_cands_seq}
\end{figure}

Only stars with a Bayesian membership probability above 90\% for any young association were selected for further consideration.\added{ This probability threshold was designed to produce similar recovery rates of 50\% for all associations in BANYAN~$\Sigma$ \bsigmap.} BANYAN~$\Sigma$ assigns each candidate member of a young association with an optimal radial velocity that maximizes its membership probability (see \bsigmalp\ for more detail). These optimal radial velocities were combined with measured kinematics to derive optimal space velocities $UVW$ that correspond to the best-matching values that can be expected for the candidate member. The candidate members with an optimal $UVW$ located at more than 5\,\kms\ or 5$\sigma$ from the core of its best-matching association kinematic model were also rejected, as this is larger than the typical spread of young association members (1--4\,\kms; \bsigmalp).\added{ This cut rejected 491 objects, which are heavily skewed toward large distances (313 of them have distances above 150\,pc).} Such candidate members might correspond to field interlopers with peculiar velocities or members of yet unknown young associations\added{, as well as more distant associations not yet included in BANYAN~$\Sigma$ (e.g., Lupus)}. These selection criteria generated a total of 1\,560 candidate members, which were cross-matched with the compilation of bona fide and candidate members compiled by \bsigma. Only the 830 stars that were not already listed in this compilation will be considered here.

The 830 candidate members were cross-matched with SIMBAD \citep{2000AAS..143...23O}, the RAVE data release 5 catalog \citep{2017AJ....153...75K} and the \cite{2007AN....328..889K} catalog to assign them radial velocity measurements. These cross-matches retrieved radial velocity with measurement errors below 10\,\kms\ for 291 stars. 67 stars were found to have a radial velocity measurement in both RAVE and other catalogs; the RAVE measurements were preferred in these cases. Only 6 of them had radial velocity measurements that differed by more than 2$\sigma$\added{; we found no evidence that these stars are binaries in the literature}. All stars with radial velocity measurements were re-analyzed with BANYAN~$\Sigma$, and the same probability and optimal $UVW$ selection criteria were used to reject an additional 122 candidate members. This resulted in a total of 539 candidate members without a radial velocity measurement, and 169 candidate members with radial velocity measurements, for a total of \nsample\ candidates which are listed in Table~\ref{tab:full}.

\begin{figure*}[p]
	\centering
	\includegraphics[width=0.95\textwidth]{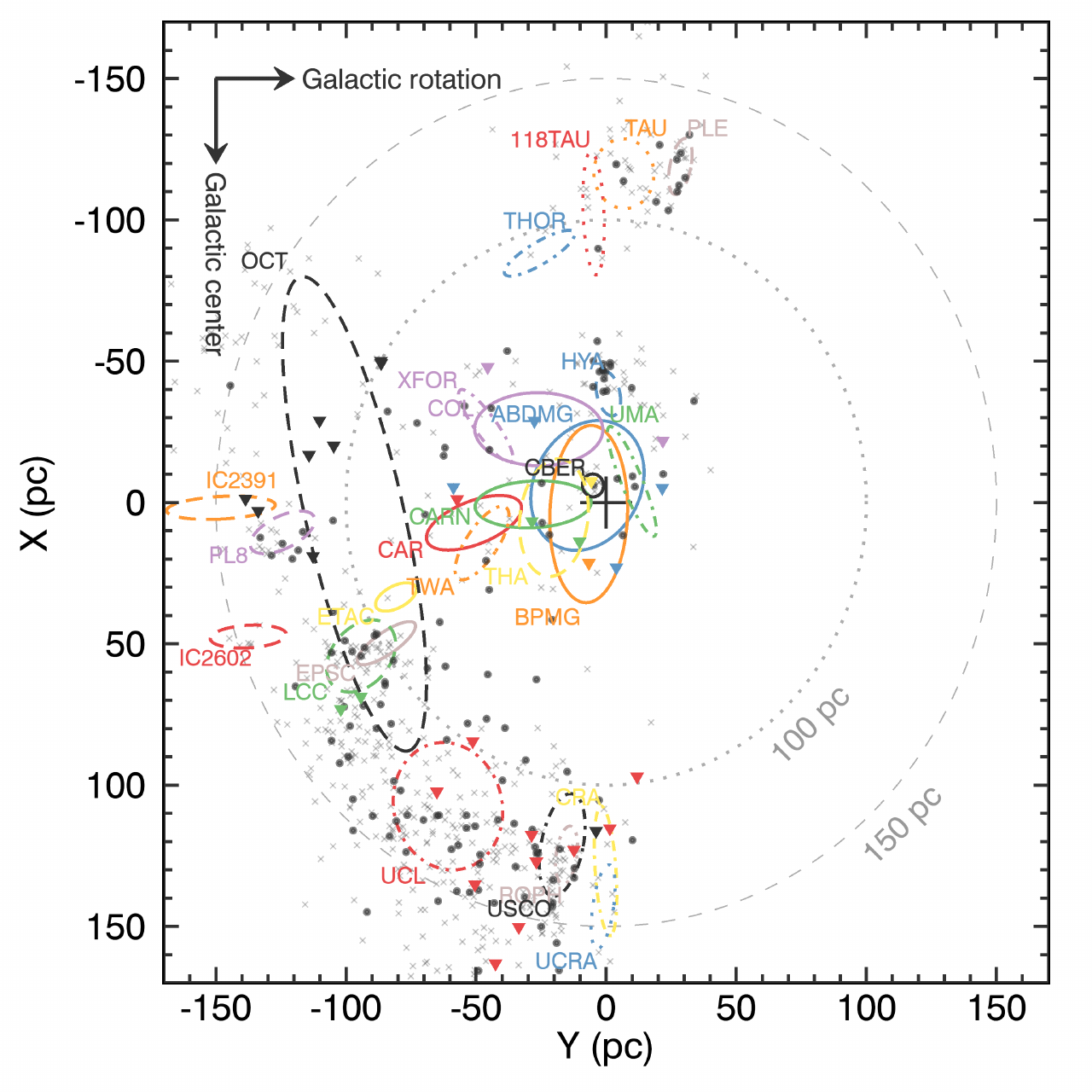}
	\caption{Galactic coordinates $XY$ of the \nsample\ candidate members uncovered in this work, compared with the 1$\sigma$ contours of the BANYAN~$\Sigma$ spatial models\added{ projected in the $XY$ plane (the full models are 6D multivariate Gaussians in $XYZUVW$ space)}. The candidates without radial velocity measurements are displayed as gray crosses, and those with full kinematics are displayed as black circles.\added{ New bona fide members identified in this work are marked as downward triangles and their color corresponds to that of their association. A handful of new UCL members are distributed toward USCO and ROPH in this 2D projection, but their $UVW$ velocities are much more consistent with UCL, which explains their classification.} See Section~\ref{sec:candidates} for more detail.}
	\label{fig:xy}
\end{figure*}

In Figure~\ref{fig:all_cands_seq}, all \nsample\ candidate members identified here are compared with the field and young color-magnitude diagrams built in Section~\ref{sec:cmd}. There are a few notable cases where stars fall well outside of the young color-magnitude sequences despite their having a Bayesian membership probability above 90\%. This indicates that they were a poor fit to even the field color-magnitude sequence. An investigation of the 6 such cases that have a radial velocity measurement (HD~145501, ROXs~43, CD--33~10685, MZ~Lup, DR~Tau and TYC~8881--551--1) reveals that they have circumstellar disks or infrared excesses that hint at a possible disk \citep{2017MNRAS.471..770M,2009ApJ...696L..84C,2003PASP..115..965E,2012ApJ...745...19K,2016ApJ...826..171G}, which likely explains their peculiar position in the color-magnitude diagram.

The Galactic coordinates $XY$ of all \nsample\ candidates are compared with the BANYAN~$\Sigma$ spatial models in Figure~\ref{fig:xy}. Ten candidates (DR~Tau and 9 others without radial velocity measurements in Table~\ref{tab:full}) are located at distances further than 200\,pc and could be members of associations not included in BANYAN~$\Sigma$.\added{ In Table~\ref{tab:ambiguous}, we list the 60 targets that have ambiguous membership in more than one young association.}

\startlongtable
\tabletypesize{\small}
\begin{deluxetable}{ll}
\tablecolumns{2}
\tablecaption{Ambiguous candidate members. \label{tab:ambiguous}}
\tablehead{\colhead{Name} & \colhead{Membership\tablenotemark{a}}}
\startdata
WOH~S~216 & ABDMG(75), BPMG(25)\\
HD~110696 & LCC(87), TWA(7), UCL(6)\\
CD--45~8100 & LCC(94), UCL(6)\\
HD~112670 & LCC(93), UCL(7)\\
HD~113975 & LCC(65), UCL(35)\\
HD~114599 & LCC(85), UCL(15)\\
HD~114788 & LCC(89), UCL(11)\\
HD~114897 & LCC(83), UCL(17)\\
HD~116116 & LCC(91), UCL(9)\\
HD~116335 & UCL(52), LCC(48)\\
CD--58~5027 & LCC(82), UCL(18)\\
HD~116553 & UCL(63), LCC(37)\\
HD~116587 & LCC(93), UCL(7)\\
HD~116649 & LCC(83), UCL(17)\\
HD~116651 & LCC(92), UCL(8)\\
HD~117353 & LCC(74), UCL(26)\\
HD~118134 & LCC(92), UCL(8)\\
TYC~8269--314--1 & LCC(63), UCL(37)\\
HD~118867 & UCL(68), LCC(32)\\
HD~119067 & UCL(86), LCC(14)\\
HD~119403 & UCL(90), LCC(10)\\
HD~119404 & UCL(76), LCC(24)\\
HD~119573 & UCL(84), LCC(16)\\
HD~120075 & UCL(89), LCC(11)\\
TYC~8664--329--1 & LCC(69), UCL(31)\\
HD~120641 & LCC(70), UCL(30)\\
HD~120795 & LCC(61), UCL(39)\\
CD--51~7806 & UCL(61), LCC(39)\\
HD~121020 & UCL(81), LCC(19)\\
HD~121191 & UCL(86), LCC(14)\\
HD~121617 & UCL(87), LCC(13)\\
HD~122414 & UCL(84), LCC(16)\\
HD~122513 & UCL(92), LCC(8)\\
HD~124746 & UCL(94), LCC(6)\\
HD~125036 & UCL(74), LCC(26)\\
HD~126181 & UCL(93), LCC(7)\\
CD--25~11037 & USCO(94), UCL(6)\\
HD~140390 & UCL(75), USCO(25)\\
HD~141960 & USCO(67), UCL(33)\\
HD~142540 & USCO(72), UCL(28)\\
HD~143069 & USCO(95), UCL(5)\\
HD~144049 & UCL(71), USCO(29)\\
HD~147754 & UCL(69), USCO(31)\\
HD~148409 & USCO(95), UCL(5)\\
HD~148606 & USCO(90), UCL(10)\\
HD~148982 & USCO(83), UCL(17)\\
HD~149514 & UCL(94), USCO(6)\\
HD~149598 & USCO(91), UCL(9)\\
TYC~6809--836--1 & USCO(92), UCL(8)\\
HD~176423 & CRA(93), UCRA(7)\\
HD~195266 & BPMG(92), CARN(8)\\
HD~115371 & TWA(82), UCL(11), LCC(6)\\
HD~117620 & LCC(94), UCL(6)\\
TYC~8273--917--1 & LCC(80), UCL(20)\\
HD~123247 & UCL(78), LCC(22)\\
Cl*~NGC~5606~VF~51 & UCL(92), LCC(8)\\
HD~142992 & UCL(74), USCO(26)\\
TYC~6801--214--1 & UCL(55), USCO(45)\\
HD~146974 & USCO(77), UCL(23)\\
TYC~6803--994--1 & USCO(90), UCL(10)\\
\enddata
\tablenotetext{a}{Possible associations are listed with their relative probabilities (\%) in parentheses. These relative probabilities add up to 100\%, and represent the relative shares of the total young association probability of each target. All targets in this table have a total young association probability above 90\%.}
\end{deluxetable}

\section{SIGNS OF YOUTH FROM THE LITERATURE}\label{sec:youth}

A literature search was performed to determine whether the candidates identified in Section~\ref{sec:candidates} were already known as candidate members of a young association, or whether they display signs of youth. A total of \news\ candidates were never identified as \deleted{such}{candidate members of young associations} in the literature, \bfidetable\ of which now have complete kinematics.\added{ These stars are listed in Table~\ref{tab:newbonafide}. There are three stars in Table~\ref{tab:newbonafide} (TYC~8881--551--1, TYC~8098--597--1 and CD--31~11053) that were identified as spectral binaries by \citep{2006AA...460..695T}. This makes them uncertain members because their orbital motion could affect the measured radial velocities, and we therefore do not draw a firm conclusion about their membership here. DR~Tau has complete kinematics consistent with TAU, but is located at a distance of $\sim$\,207\,pc, significantly larger than its other members ($\sim$\,120\,pc). It is possible that TAU is spatially much larger than previously thought, but we do not draw any firm conclusions on the membership of DR~Tau here. Similarly, CD--33~10685 (at a distance of $\sim$\,141\,pc) has complete kinematics consistent with UCL, but \cite{2008ApJS..177..551M} classify it as a member of Lupus, which is too distant to have been included in the BANYAN~$\Sigma$ models. We will therefore wait for the future inclusion of Lupus in the BANYAN~$\Sigma$ models before drawing a conclusion on its membership.}

\begin{figure}
	\centering
	\includegraphics[width=0.485\textwidth]{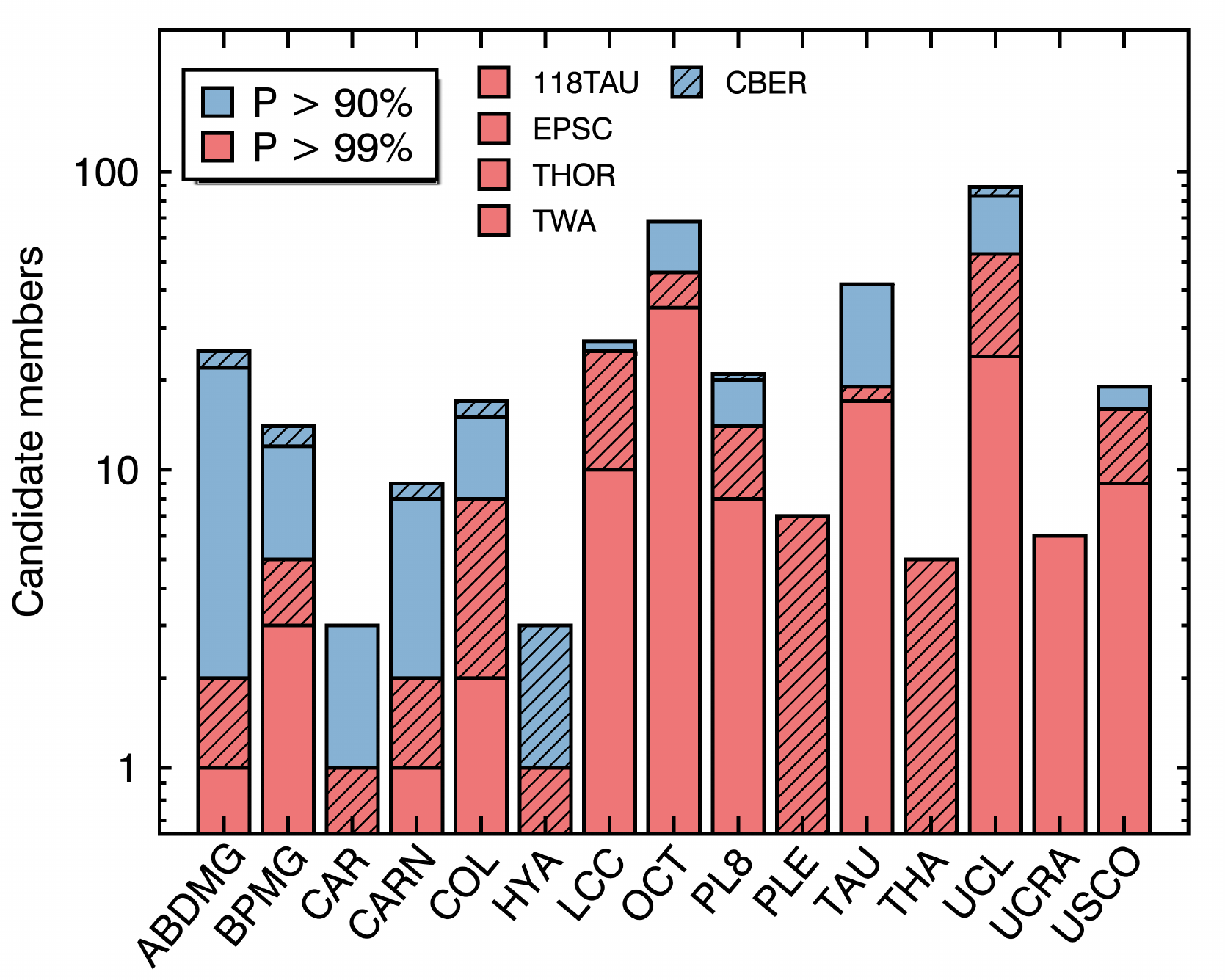}
	\caption{Population histogram of new candidate members of each young association identified in this work. Candidates with diagonal hashing have radial velocity measurements. A few associations are displayed separately\added{ at the top of the figure} because they each have only one candidate member. See Section~\ref{sec:youth} for more detail.}
	\label{fig:hist_new_cand}
\end{figure}

\begin{figure}
	\centering
	\includegraphics[width=0.485\textwidth]{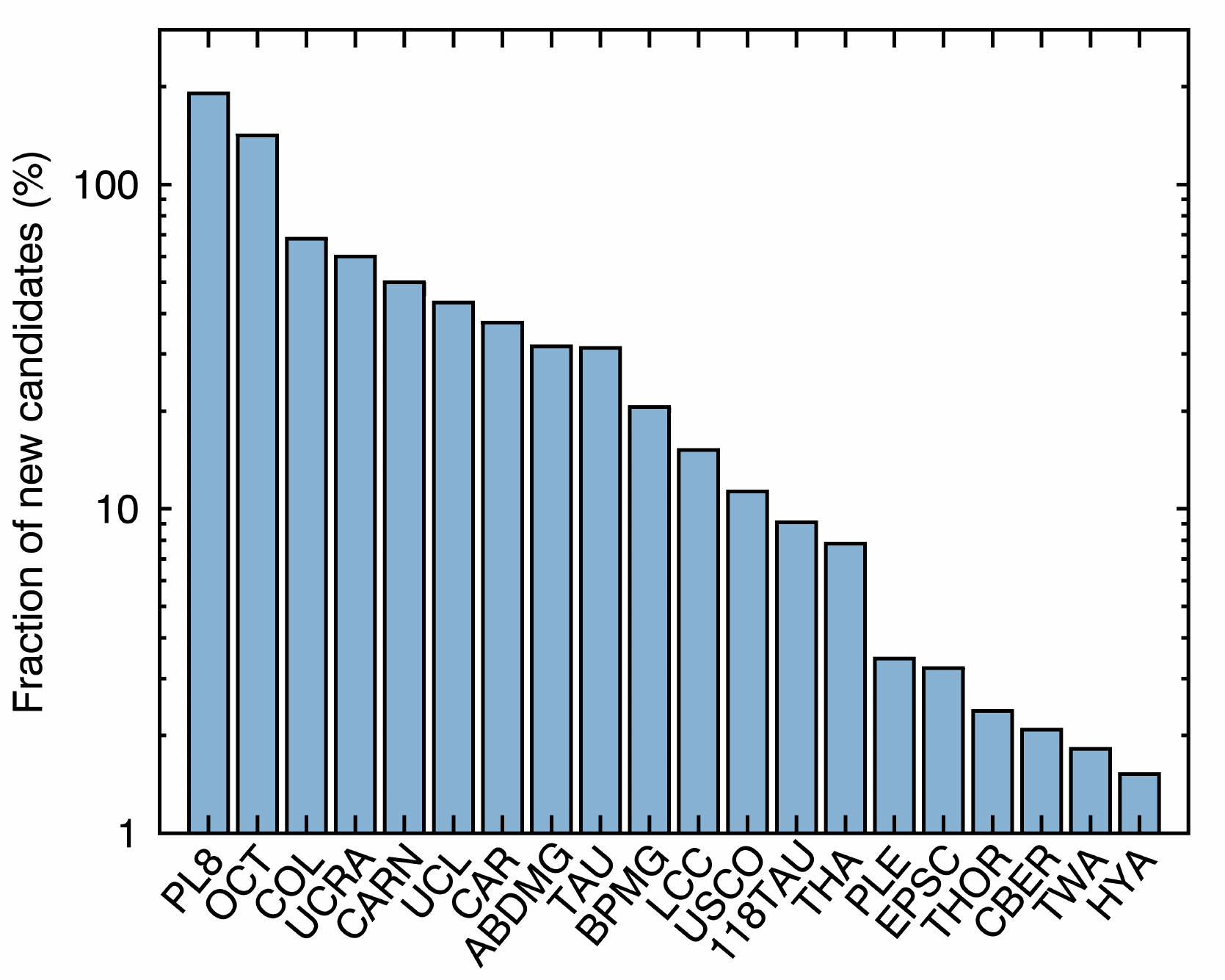}
	\caption{Fractional population histogram of new candidate members of each young association identified in this work, divided by the number of known bona fide members. Some associations such as PL8 and OCT have not been extensively studied in the literature. See Section~\ref{sec:youth} for more detail.}
	\label{fig:hist_frac_new_cand}
\end{figure}

We found that \missedbfs\ objects were already known as bona fide members with full kinematics and were either overlooked or excluded in the \bsigma\ compilation of bona fide members. For example, some members of HYA identified by \cite{2017AA...601A..19G} were not clearly identified as cluster or stream members, and some ABDMG members identified by \citep{2004ARAA..42..685Z} had a ``Questionable membership flag''. We consider that their high membership probability calculated in this work warrants considering them as bona fide members, as it demonstrates that they have kinematics consistent with other unambiguous members. A histogram of the associations in which new candidate members were identified here is displayed in Figure~\ref{fig:hist_new_cand}.\added{ In Figure~\ref{fig:hist_frac_new_cand}, we display the fractional number of new candidate members identified here compared to the number of currently known bona fide members. Some associations such as PL8 and OCT were not extensively studied in the literature, and as a consequence our sample will make a significant contribution to their number of known members upon full confirmation of their kinematics.}

\deleted{\confirmedcmstxt\ }\added{\confirmedplusreclassedcmstxt\ }others objects were already known as candidate members, and the addition of a \gaia\ parallaxe or a newly compiled radial velocity confirms their membership\added{ in \confirmedcmstxt\ cases,} or revises it to a different association in \reclassedconfirmed\ cases. One other star (EX~Cet) that we identify as a candidate member of $\beta$PMG was classified as a member of the Hercules Lyra (HLY) association by \cite{2006ApJ...643.1160L}, but this association was not included in BANYAN~$\Sigma$ because it is likely composed of non-coeval stars \citep{2015IAUS..314...21M}. Given that EX~Cet is located at 4\,\kms\ from the core of the BANYAN~$\Sigma$ kinematic model of $\beta$PMG, we do not draw any conclusion on its membership here.\deleted{ \bfidetabletxt\ candidates with radial velocities do not seem to have been identified as candidate members in the literature. These stars are listed in Table~\ref{tab:newbonafide}.}

\startlongtable
\tabletypesize{\small}
\begin{deluxetable*}{llccccclcc}
\tablecolumns{10}
\tablecaption{Candidate members with full kinematics. \label{tab:newbonafide}}
\tablehead{\colhead{Name} & \colhead{Spectral} & \colhead{Isochronal} & \colhead{$G-J$} & \colhead{$NUV-G$} & \colhead{X-Ray} & \colhead{Li EW} & \colhead{Signs of} & \colhead{Bona fide} & \colhead{Ref.\tablenotemark{d}}\\
\colhead{} & \colhead{Type\tablenotemark{a}} & \colhead{Age\tablenotemark{b} (Myr)} & \colhead{(mag)} & \colhead{(mag)} & \colhead{HR1} & \colhead{(m\AA)} & \colhead{youth\tablenotemark{c}} & \colhead{member} & \colhead{}
}
\startdata
\sidehead{\textbf{ABDMG}}
CD--26~1643 & F9V & $\cdots$ & $0.9$ & $5.0$ & $-0.15$ & $\cdots$ & X & Y & 1,--\\
HD~61518 & F5V & $\cdots$ & $0.8$ & $\cdots$ & $-0.23$ & $\cdots$ & IR & Y & 2,--\\
HD~147512 & G8/K0V & $\cdots$ & $1.0$ & $6.3$ & $\cdots$ & $\cdots$ & IR & Y & 3,--\\
HD~221239 & K2.5V & $\cdots$ & $1.3$ & $7.4$ & $-1.00$ & $\cdots$ & UV & Y & 4,--\\
\sidehead{\textbf{BPMG}}
TYC~8098--597--1\tablenotemark{e} & K3V & $\cdots$ & $2.0$ & $\cdots$ & $0.16$ & $25$ & X & ? & 5,6\\
HD~207043 & G5V & $\cdots$ & $1.0$ & $5.9$ & $\cdots$ & $\cdots$ & IR & Y & 7,--\\
\sidehead{\textbf{CAR}}
HD~37402 & F6V & $\cdots$ & $0.8$ & $4.8$ & $-0.17$ & $110$ & Li,IR & Y & 8,9\\
\sidehead{\textbf{CARN}}
S1*~329 & K7V(ke) & $\cdots$ & $2.1$ & $8.5$ & $0.47$ & $\cdots$ & X,UV & Y & 7,--\\
L~106--104 & M3 & $\cdots$ & $2.2$ & $\cdots$ & $\cdots$ & $\cdots$ & IR,Ca & Y & 10,--\\
\sidehead{\textbf{COL}}
HD~29329 & F7V & $\cdots$ & $0.8$ & $\cdots$ & $-0.01$ & $88$ & X,IR & Y & 11,11\\
TYC~8881--551--1\tablenotemark{e} & K0IV/V & $\cdots$ & $1.8$ & $6.1$ & $-0.22$ & $\cdots$ & UV,Sp & ? & 12,--\\
V*~AI~Lep & G2V & $25_{-5}^{+3}$ & $1.1$ & $\cdots$ & $0.00$ & $213$ & X,Li,IS & Y & 11,11\\
\sidehead{\textbf{LCC}}
TYC~8649--1758--1 & (K2) & $13_{-4}^{+5}$ & $1.5$ & $\cdots$ & $\cdots$ & $\cdots$ & IR,Ca,IS & Y & --,--\\
TYC~8653--1049--1 & (G8) & $28_{-6}^{+3}$ & $1.3$ & $\cdots$ & $\cdots$ & $\cdots$ & IR,Ca,IS & Y & --,--\\
\sidehead{\textbf{OCT}}
TYC~7053--832--1 & (G6) & $\cdots$ & $1.2$ & $5.8$ & $\cdots$ & $\cdots$ & UV,IR,Ca & Y & --,--\\
HD~35212 & F5V & $\cdots$ & $0.8$ & $\cdots$ & $-0.27$ & $\cdots$ & IR & Y & 2,--\\
HD~275012 & G5 & $\cdots$ & $1.2$ & $\cdots$ & $-0.01$ & $\cdots$ & X,IR,Ca & Y & 13,--\\
CD--35~2433 & (G2) & $\cdots$ & $0.9$ & $\cdots$ & $-0.02$ & $\cdots$ & X,IR & Y & --,--\\
HD~42122 & F7/G0V & $18_{-4}^{+2}$ & $0.8$ & $4.7$ & $\cdots$ & $\cdots$ & IR,IS & Y & 8,--\\
TYC~8534--1810--1 & (G7) & $45_{-9}^{+70}$ & $1.2$ & $\cdots$ & $0.11$ & $\cdots$ & X,IR,Ca,IS & Y & --,--\\
TYC~8895--112--1 & (G4) & $\cdots$ & $1.0$ & $\cdots$ & $-0.32$ & $\cdots$ & IR,Ca & Y & --,--\\
TYC~8104--898--1 & (G1) & $\cdots$ & $0.9$ & $\cdots$ & $\cdots$ & $\cdots$ & $\cdots$ & N & --,--\\
TYC~9178--1390--1 & (K1) & $40 \pm 10$ & $1.4$ & $\cdots$ & $\cdots$ & $\cdots$ & IR,IS & Y & --,--\\
TYC~9341--1233--1 & (G9) & $\cdots$ & $1.3$ & $8.0$ & $\cdots$ & $\cdots$ & $\cdots$ & N & --,--\\
\sidehead{\textbf{TAU}}
HD~284659 & A2 & $8_{-2}^{+1}$ & $0.2$ & $\cdots$ & $\cdots$ & $\cdots$ & IS & ? & 13,--\\
\sidehead{\textbf{THA}}
HD~10863 & F0IV & $\leq 1000$ & $0.6$ & $5.5$ & $\cdots$ & $\cdots$ & IS,Sp & Y & 14,--\\
\sidehead{\textbf{UCL}}
CD--31~11053\tablenotemark{e} & K3Ve & $5 \pm 1$ & $1.8$ & $\cdots$ & $0.68$ & $470$ & X,Li,IR,IS & ? & 6,6\\
TYC~7295--853--1 & (G7) & $16_{-3}^{+2}$ & $1.2$ & $\cdots$ & $\cdots$ & $\cdots$ & IR,Ca,IS & Y & --,--\\
TYC~7296--1194--1 & (K0) & $22_{-5}^{+3}$ & $1.3$ & $\cdots$ & $\cdots$ & $\cdots$ & Ca,IS & Y & --,--\\
TYC~7353--768--1 & G8V & $18_{-5}^{+2}$ & $1.1$ & $\cdots$ & $1.00$ & $270$ & X,Li,IR,IS & Y & 6,6\\
HD~321958 & G9V & $14_{-3}^{+2}$ & $1.2$ & $\cdots$ & $0.51$ & $275$ & X,Li,IR,IS & Y & 6,6\\
TYC~8332--2024--1 & K5Ve & $\cdots$ & $1.7$ & $\cdots$ & $0.03$ & $480$ & X,Li & Y & 6,6\\
V*~V991~Sco & G6/8 & $9 \pm 1$ & $1.1$ & $\cdots$ & $-0.05$ & $\cdots$ & X,IR,IS & Y & 15,--\\
CD--23~13197 & K0IV(e) & $13_{-4}^{+2}$ & $1.3$ & $\cdots$ & $0.52$ & $360$ & X,Li,IR,IS,Sp & Y & 6,6\\
HD~317637 & K2V & $13_{-3}^{+2}$ & $1.4$ & $\cdots$ & $0.03$ & $390$ & X,Li,IR,IS & Y & 6,6\\
BD--18~4557 & K2IV(e) & $9_{-2}^{+1}$ & $1.5$ & $\cdots$ & $0.18$ & $420$ & X,Li,IR,IS,Sp & Y & 6,6\\
CD--43~11887 & G9V & $9_{-3}^{+2}$ & $1.3$ & $\cdots$ & $0.13$ & $350$ & X,Li,IR,IS & Y & 6,6\\
\sidehead{\textbf{USCO}}
CD--25~11942 & K0IV & $6.3_{-0.7}^{+0.8}$ & $1.4$ & $\cdots$ & $0.01$ & $310$ & X,Li,IS,Sp & Y & 6,6\\
\enddata
\tablenotetext{a}{Spectral types in parentheses were estimated using the $G - J$ color with the spectral type--color relations of \cite{2013ApJS..208....9P}, see also \url{http://www.pas.rochester.edu/~emamajek/EEM_dwarf_UBVIJHK_colors_Teff.txt}.}
\tablenotetext{b}{Pre-main sequence ages derived from MIST isochrones in absolute $G$ versus $G-J$. See Section~\ref{sec:youth} for more detail. The color-magnitude positions of HD~284659, HD~42122 and TYC~8534--1810--1 are also consistent with respective post-main sequence ages of $560_{-100}^{+70}$\,Myr, $4.5_{-0.9}^{+0.5}$\,Gyr and $10_{-6}^{+4}$\,Gyr. }
\tablenotetext{c}{Signs of youth compiled from the literature. See Section~\ref{sec:youth} for more detail. X: X-ray emission with HR1 $\geq$ -0.15; UV: Galex $NUV-G$ versus $G-J$ consitent with youth; Li: Lithium absorption above 100\,m\AA; E: Mid-infrared excess; Lm: Luminosity class consistent with youth; Is: Young isochronal age consistent with; Ca: \ion{Ca}{2} infrared triplet age consistent with proposed association.}
\tablenotetext{d}{References for: (1) spectral type and (2) Li absorption.}
\tablenotetext{e}{Spectral binary \citep{2006AA...460..695T}.}
\tablerefs{(1)~\citealt{1964PLPla..28....1J}; (2)~\citealt{1978mcts.book.....H}; (3)~\citealt{1999MSS...C05....0H}; (4)~\citealt{2003AJ....126.2048G}; (5)~\citealt{2001KFNT...17..409K}; (6)~\citealt{2006AA...460..695T}; (7)~\citealt{2006AJ....132..161G}; (8)~\citealt{1975mcts.book.....H}; (9)~\citealt{2013MNRAS.435.1376M}; (10)~\citealt{1996AAS..116..301R}; (11)~\citealt{2007AJ....133.2524W}; (12)~\citealt{2004AA...418..989N}; (13)~\citealt{1995AAS..110..367N}; (14)~\citealt{2014MNRAS.443.2815P}; (15)~\citealt{1982mcts.book.....H}.}
\end{deluxetable*}

\begin{figure}
	\centering
	\includegraphics[width=0.485\textwidth]{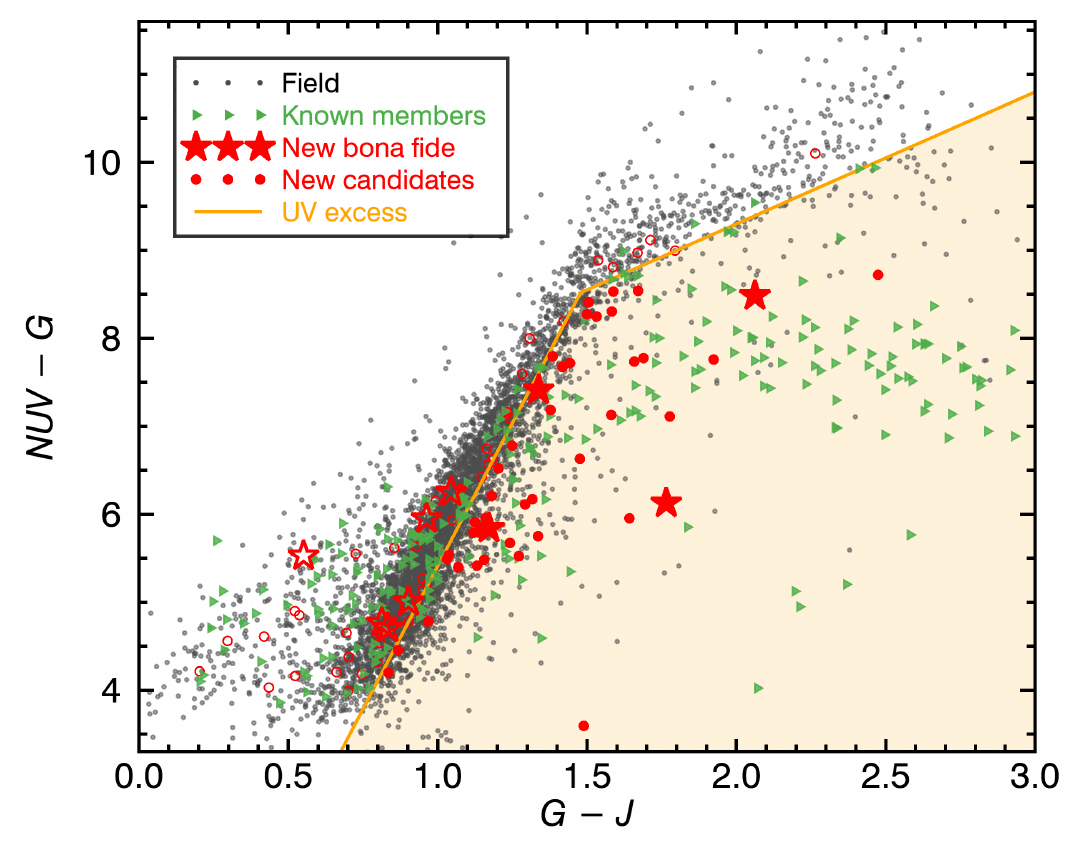}
	\caption{\emph{GALEX} $NUV$ excess for field stars (black dots), known bona fide members of young associations (green triangles) and the new candidates (red circles) and bona fide members (red stars) identified here. The criterion for $NUV$ excess defined in Equation~\eqref{eqn:uv} is displayed in orange, and the new candidates identified in this work that are selected by this criterion are displayed with filled symbols. See Section~\ref{sec:youth} for more detail.}
	\label{fig:nuv}
\end{figure}

The $G-J$ versus $NUV-G$ colors of the stars with an entry in the data release 5 of \emph{GALEX} \citep{2005ApJ...619L...1M} were compared with the field and young sequences and are displayed in Figure~\ref{fig:nuv}.\added{ A total of 133 stars in our sample have an entry in \emph{GALEX}; the main reason for this incompleteness is the poor coverage of \emph{GALEX} at Galactic latitudes South of $l = 30$\textdegree.} This figure is similar to Figure~2 of \cite{2011ApJ...727...62R}, but uses \emph{Gaia} colors instead of those based on Tycho $V$. The field sequence was built from a combination of the field sample of \cite{2011ApJ...727...62R}, complemented with all stars in the sample of \cite{2017AJ....153..257O}, from which all known young stars were removed, as well as any star with a BANYAN~$\Sigma$ probability above 1\% of belonging to a known young association. Any groups of stars in the \cite{2017AJ....153..257O} sample that display signs of youth as a population (see J.~K.~Faherty et al., in preparation) were also excluded from our field sample. The sample of young associations compiled by \bsigma\ are compared with the field $NUV$ sequence in Figure~\ref{fig:nuv}, which allowed us to derive a simple criterion that delimitates regions dominated by young stars:
\begin{align}
	NUV - G &< 6.5\left(G-J\right) - 1.09 \mbox{  if $G-J < 1.476$},\notag\\
	NUV - G &< 1.5\left(G-J\right) - 6.30 \mbox{  if $G-J \geq 1.476$},\label{eqn:uv}
\end{align}

This criterion is especially efficient at identifying the $NUV$ excess of G-type stars or later ($G - J > 1$). For earlier-type stars, the young and field sequences become gradually harder to distinguish and this criterion will fail to identify most young stars.

\begin{figure}
	\centering
	\includegraphics[width=0.485\textwidth]{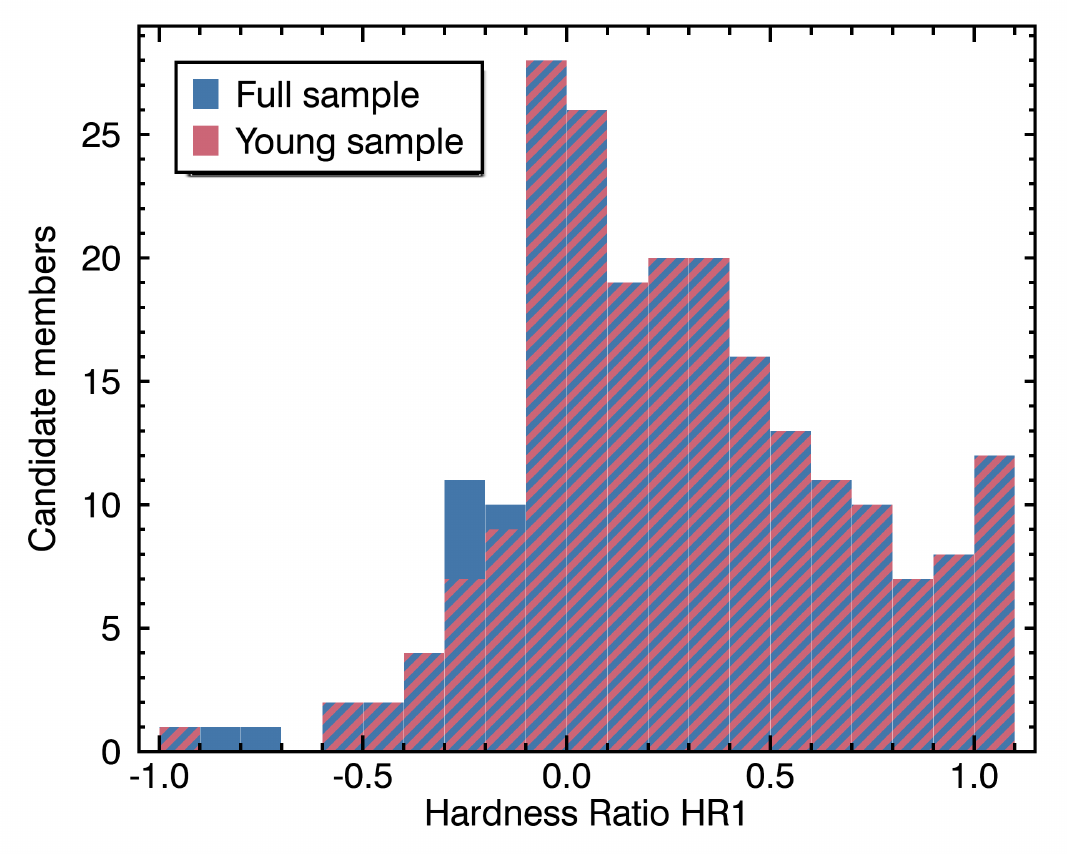}
	\caption{Histogram of \emph{ROSAT} hardness ratios HR1 for the full and young samples. The full sample is skewed toward high HR1 values as expected, and most stars  with X-ray detections are young. See Section~\ref{sec:youth} for more detail.}
	\label{fig:hrhist}
\end{figure}
\begin{figure}
	\centering
	\includegraphics[width=0.485\textwidth]{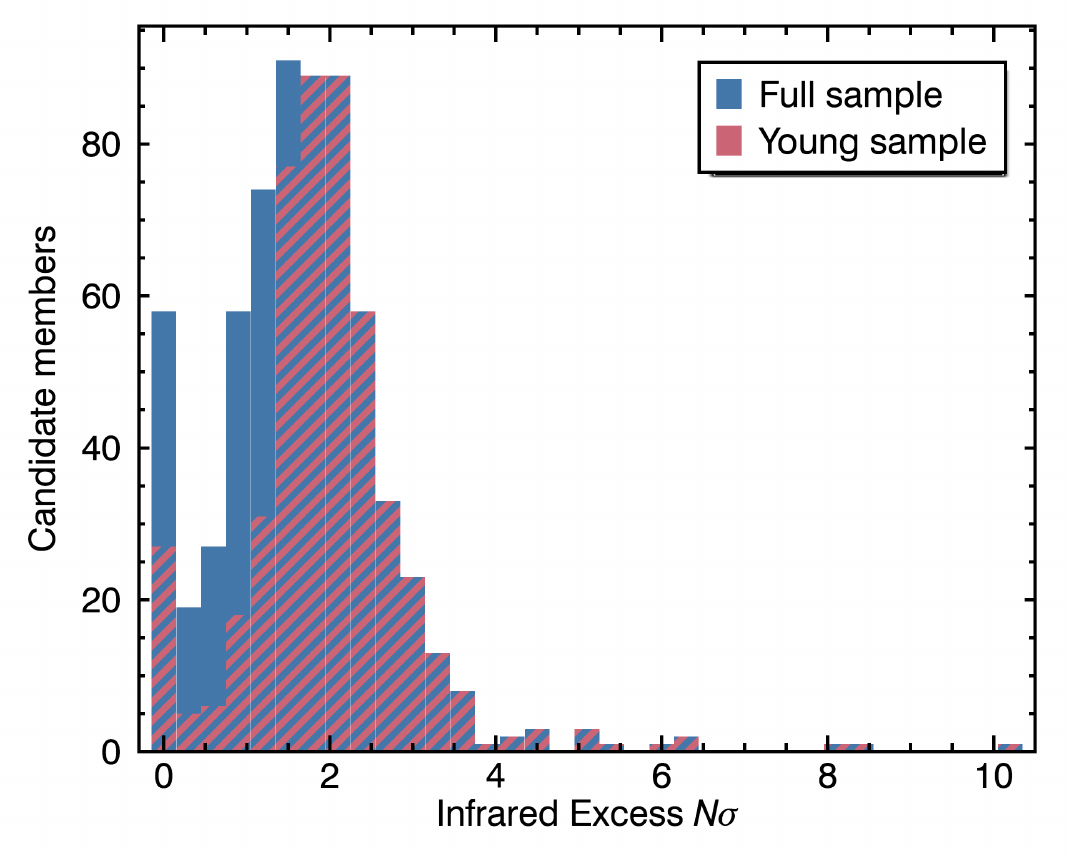}
	\caption{Histogram of the statistical significance $N\sigma$ of infrared excess for the full and young samples. A large fraction of stars in the full sample has a significant excess, and this is even more true of the young sample. See Section~\ref{sec:youth} for more detail.}
	\label{fig:irhist}
\end{figure}

The candidate members identified here were similarly cross-matched with the second \emph{ROSAT} all-sky survey catalog \citep{2016AA...588A.103B} and the XMM-Newton slew survey Source Catalogue v2.0 \citep{2018yCat.9053....0X}.\added{ A total of 222 stars in our sample have an entry in either one of these catalogs. The main limitation of this cross-match is the limited \emph{ROSAT} sensitivity, which does not allow it to detect the furthest and/or lowest-mass stars in our sample.} All stars with a HR1 hardness ratio above -0.15, consistent with the observed distribution of young $\beta$PMG, TWA and THA members in \cite{2003ApJ...585..878K}, were flagged as likely young.\added{ The distribution of HR1 values are displayed in Figure~\ref{fig:hrhist}.} Stars with a mid-infrared excess detected at a $\geq$\,1.5$\sigma$ statistical significance in the \cite{2017MNRAS.471..770M} catalog were also flagged as likely young.\added{ \cite{2017MNRAS.471..770M} analyzed most of the \gaia\ sample, but some of the stars could not be cross-matched to older catalogs because they ignored their proper motion, and they used a more conservative cut on the distance precision (they required a statistical significance above 2.4$\sigma$ versus our 2$\sigma$ requirement), which resulted in only 656 stars in our sample having such a mid-infrared excess measurement.\added{ These measurements are displayed in Figure~\ref{fig:irhist}.}} \cite{2017ApJ...835...61Z} provide age constraints with a precision of $\sim$\,50\% for RAVE survey stars based on the \ion{Ca}{2} infrared triplet chromospheric activity indicator. There are 17 stars in our sample that have such activity-based ages consistent with their respective association. Lithium equivalent widths were also reported by \cite{2006AA...460..695T}, \cite{2007AJ....133.2524W}, \cite{2009AA...508..833D}, \cite{2013MNRAS.435.1376M} and \cite{2016MNRAS.461..794P} for 114 stars in our sample. Measurements of lithium equivalent widths above 100\,m\AA\ were adopted as a sign of youth (e.g., see \citealp{2013MNRAS.435.1376M}). A weaker lithium line is not necessarily inconsistent with youth, but does not provide a strong indication of youth. All youth indicators based on $NUV$, X-ray, lithium, infrared excess or the \ion{Ca}{2} infrared triplet are reported in Tables~\ref{tab:newbonafide} and \ref{tab:full}.

\begin{figure*}
	\centering
	\includegraphics[width=0.95\textwidth]{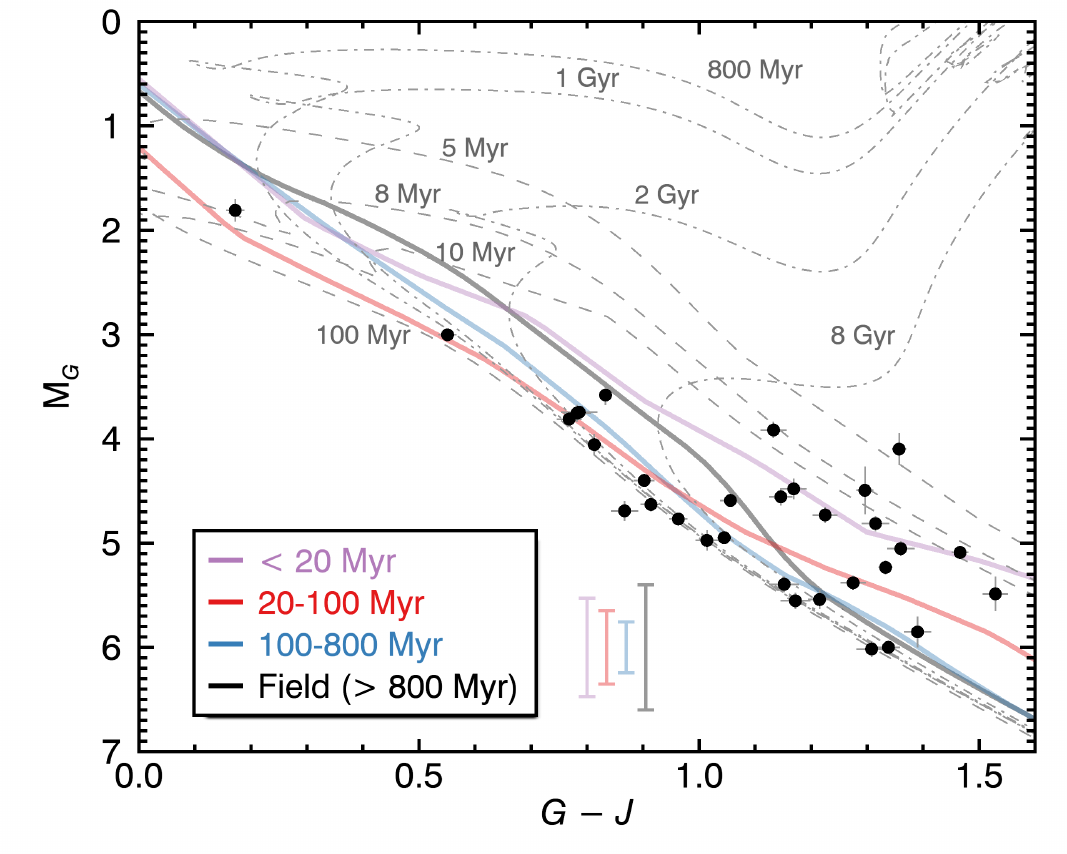}
	\caption{MIST isochrones compared to empirical color-magnitude sequences (thick colored lines) and objects from Table~\ref{tab:newbonafide} (black circles).\added{ There are 18 stars in this figure for which the isochrones do not provide a useful age constraint; those have no age measurement in Table~\ref{tab:newbonafide}.} The median scatter of each empirical sequence is indicated in the lower portion of the figure. MIST isochrones younger (older) than 500\,Myr are represented with dashed (dash-dotted) lines for clarity. See Section~\ref{sec:youth} for more detail.}
	\label{fig:isoc_bf}
\end{figure*}

We used the MIST solar-metallicity model tracks of \cite{2016ApJ...823..102C} to determine isochronal ages for the \bfidetable\ new candidate members with complete kinematics. To do so, we calculated the minimum $N\sigma$ distance of the star to each isochrone in absolue \emph{Gaia} $M_G$ versus \emph{Gaia}--2MASS $G-J$ to build a probability density function as a function of age. The probability density functions were then visually inspected, and were classified in one of three categories to determine whether (1) they do not provide a significant age constraint; (2) they provide a unimodal age constraint, or (3) they provide a bimodal age constraint (i.e., a pre-main sequence age and a post-main sequence age). The resulting pre-main sequence ages are reported in Table~\ref{tab:newbonafide}, and corroborate membership in all cases but one; the pre-main sequence isochronal age of the A2 TAU candidate HD~284659 is $8_{-2}^{+1}$\,Myr, which is significantly older than the estimated age of TAU (1--2\,Myr; \citealt{1995ApJS..101..117K}). It remains unclear whether HD~284659 is a 1--2\,Myr member of TAU (in which case the MIST tracks systematically over-estimate the age in this very young regime), or if it is part of an older sub-group with similar kinematics to TAU that is not considered in BANYAN~$\Sigma$. HD~284659 is located spatially within the distribution of TAU members (at 5.8\,pc, or 0.6$\sigma$ from the core of the BANYAN~$\Sigma$ spatial model), and at 3.9\,\kms\ (or 1$\sigma$) from the core of the BANYAN~$\Sigma$ kinematic model. Finding more 6--9\,Myr objects in the vicinity of HD~284659, or calibrating the MIST tracks for 1--2\,Myr A-type stars using more empirical data, would help in clarifying the membership of HD~284659. Alternatively, it is possible that HD~284659 is instead a $560_{-100}^{+70}$\,Myr-old star that is starting to depart from the main sequence and that happens to share the spatial position and kinematics of TAU by pure chance. Figure~\ref{fig:isoc_bf} displays the MIST solar-metallicity tracks compared with stars of Table~\ref{tab:newbonafide} and empirical color-magnitude sequences derived in Section~\ref{sec:cmd}.

A total of \bfide\ candidate members with full kinematics have at least one sign of youth that is consistent with their kinematic membership, and we therefore propose them as new bona fide members of their respective young associations. The OCT candidate member TYC~8104--898--1 has no literature information that allows us to put a constraint on its age, and its position in a $M_G$ versus $G-J$ color-magnitude diagram is slightly under-luminous compared to all solar-metallicity MIST tracks. The RAVE data release 5 catalog \citep{2017AJ....153...75K} provides a sub-solar metallicity measurement of [m/H]$ = -0.26 \pm 0.09$\,dex for this star, which may explain its peculiar absolute magnitude. We therefore reject TYC~8104--898--1 as a candidate member of OCT. TYC~9341--1233--1 is also rejected because its \emph{GALEX}--\emph{Gaia} $NUV-G$ color is consistent with field stars, and no conclusion is drawn on the TAU membership of HD~284659 because of its inconsistent (but likely young) isochronal age.

\added{Seven stars in our sample have signs of youth but are rejected as candidate members of young associations. These stars could be slightly older and have formed in groups that are now completely dissolved; scattered members of known associations; or members of associations not yet discovered. Similarly, several young brown dwarfs without a clear origin were identified in previous work (e.g., \citealt{2015ApJS..219...33G,2016ApJS..225...10F}). The upcoming of \emph{Gaia}~DR2 will help understanding the origin of these objects.}

\section{CONCLUSIONS}\label{sec:conclusion}

We used the BANYAN~$\Sigma$ tool in combination with \gaia\ to identify \bfide\ new F0--M3 bona fide members of nearby young associations and \newcms\ additional new candidate members.\added{These new candidate members include HD~121191, a new A-type candidate member of UCL which is known to have a significant infrared excess \citep{2017MNRAS.471..770M,2013ApJ...778...12M}, and 3 other infrared excess candidate members of associations older than $\sim$\,10\,Myr (TYC~8602--718--1 in $\beta$PMG; CD--60~2373 in PL8; TYC~8602--718--1 in CAR). The discovery of accretion disks around these slightly older stars would be valuable to understand their lifetimes, as only few such examples are currently known (e.g., \citealp{2016ApJ...832...50B,2016ApJ...830L..28S,2018MNRAS.476.3290M}). Five new candidate members of CARN and ABDMG are located within 30\,pc of the Sun (the nearest one, HD~19819, is a candidate member of CARN at $\sim$\,22\,pc), and 26 objects in our sample are new A-type candidate members of young associations. These stars will be particularly valuable for direct-imaging searches of exoplanets, as proximity makes it possible to detect companions at smaller separations, and massive stars are known to have a larger occurrence of giant exoplanet companions which are easier to detect \citep{2017AA...603A..54L}. Furthermore, 26 new candidate members (21 with signs of youth) of OCT, TAU, UCL and USCO are located at spatial distances above 150\,pc, making them $> 5$\,$\sigma$ outliers to the spatial model of their respective association, while displaying consistent kinematics. This hints that these associations may extend to larger distances not yet explored due to the paucity of parallax measurements previously available beyond 150\,pc. The new candidates presented here have the potential to almost double the number of members in associations that were not extensively studied in the literature such as PL8 and OCT.}

This work hints that the upcoming \emph{Gaia}--DR2 will allow us to uncover many new young stars, complete their kinematic picture and investigate the initial mass function of several young associations down to the regime of low-mass stars. The identification of new young stars and their assignment of accurate ages will be useful to build standard stellar populations, and will provide valuable targets for direct-imaging searches of exoplanets.

\begin{changemargin}{720pt}
\startlongtable
\tabletypesize{\tiny}
\begin{longrotatetable}
\global\pdfpageattr\expandafter{\the\pdfpageattr/Rotate 90}
\begin{deluxetable*}{lllllr@{\;$\pm$\;}lr@{\;$\pm$\;}lr@{\;$\pm$\;}lcccllll}
\tablecolumns{18}
\tablecaption{Candidate members recovered in this work. \label{tab:full}}
\tablehead{\colhead{Name} & \colhead{Assoc.} & \colhead{Spectral} & \colhead{R.A.} & \colhead{Decl.} & \multicolumn{2}{c}{$\mu_\alpha\cos\delta$}  & \multicolumn{2}{c}{$\mu_\delta$} & \multicolumn{2}{c}{Distance} & \colhead{Rad. Vel.} & \colhead{HR1} & \colhead{$NUV$} & \colhead{IR$_{\rm ex}$} & \colhead{Known} & \colhead{Youth\tablenotemark{b}} & \colhead{Ref.\tablenotemark{c}}\\
\colhead{} & \colhead{} & \colhead{Type\tablenotemark{a}} & \colhead{(hh:mm:ss)} & \colhead{(dd:mm:ss)} & \multicolumn{2}{c}{(\masyr)} & \multicolumn{2}{c}{(\masyr)} & \multicolumn{2}{c}{(pc)} & \colhead{(\kms)} & \colhead{} & \colhead{(mag)} & \colhead{$N\sigma$} & \colhead{Group} & \colhead{} & \colhead{}
}
\startdata
\sidehead{\textbf{New Bona Fide Members}}
CD--26~1643 & ABDMG & F9V & 04:21:33.518 & --25:43:10.14 & $42.74$ & $0.31$ & $-55.46$ & $0.33$ & $54.84$ & $0.85$ & $23 \pm 2$ & $-0.15$ & $13.099 \pm 0.004$ & $0.0$ & $\cdots$ & X & 1,2,--,--\\
HD~61518 & ABDMG & F5V & 07:37:02.743 & --52:21:56.94 & $-31.93$ & $0.05$ & $6.13$ & $0.06$ & $60.95$ & $0.84$ & $31 \pm 1$ & $-0.23$ & $\cdots$ & $2.7$ & $\cdots$ & E & 3,4,--,--\\
HD~147512 & ABDMG & G8/K0V & 16:22:32.642 & --04:14:59.94 & $-78.92$ & $0.05$ & $-176.73$ & $0.03$ & $27.03$ & $0.24$ & $-19.9 \pm 0.7$ & $\cdots$ & $13.358 \pm 0.004$ & $1.5$ & $\cdots$ & E & 5,5,5,--\\
HD~221239 & ABDMG & K2.5V & 23:30:13.963 & +31:42:19.02 & $191.28$ & $0.06$ & $-130.31$ & $0.05$ & $25.11$ & $0.16$ & $-18 \pm 4$ & $-1.00$ & $15.41 \pm 0.01$ & $0.0$ & $\cdots$ & UV & 6,4,--,--\\
HD~37402 & CAR & F6V & 05:34:26.199 & --60:06:14.59 & $25.76$ & $0.04$ & $32.20$ & $0.05$ & $68.2$ & $3.0$ & $23.7 \pm 0.5$ & $-0.17$ & $12.987 \pm 0.005$ & $3.0$ & $\cdots$ & Li,E & 9,4,10,--\\
S1$^{*}$~329 & CARN & K7V(ke) & 08:41:32.326 & --68:25:37.77 & $-113.29$ & $0.09$ & $164.47$ & $0.09$ & $30.45$ & $0.49$ & $10 \pm 2$ & $0.47$ & $18.77 \pm 0.08$ & $1.4$ & $\cdots$ & X,UV & 11,12,13,--\\
L~106--104 & CARN & M3 & 22:46:45.574 & --63:18:06.00 & $198.01$ & $0.14$ & $-113.86$ & $0.14$ & $26.00$ & $0.23$ & $-4 \pm 4$ & $\cdots$ & $\cdots$ & $2.2$ & $\cdots$ & E,Ca & 14,7,--,--\\
HD~29329 & COL & F7V & 04:46:00.903 & +76:36:37.73 & $75.59$ & $0.02$ & $-135.35$ & $0.03$ & $32.75$ & $0.34$ & $-9.8 \pm 0.8$ & $-0.01$ & $\cdots$ & $2.6$ & $\cdots$ & X,E & --,7,--,--\\
\enddata
\end{deluxetable*}
\end{longrotatetable}
\global\pdfpageattr\expandafter{\the\pdfpageattr/Rotate 0}
\end{changemargin}

\acknowledgments

\added{We thank the anonymous referee for useful comments. }We thank Dustin Lang for providing the \gaia--2MASS cross-match data, David Rodriguez from providing part of the data used to build the field sequence in Figure~\ref{fig:nuv}, and Eric E. Mamajek for useful comments. This research made use of: the SIMBAD database and VizieR catalog access tool, operated at the Centre de Donn\'ees astronomiques de Strasbourg, France \citep{2000AAS..143...23O}; data products from the Two Micron All Sky Survey (\emph{2MASS}; \citealp{2006AJ....131.1163S}), which is a joint project of the University of Massachusetts and the Infrared Processing and Analysis Center (IPAC)/California Institute of Technology (Caltech), funded by the National Aeronautics and Space Administration (NASA) and the National Science Foundation \citep{2006AJ....131.1163S}; data products from the \emph{Wide-field Infrared Survey Explorer} (\emph{WISE}; and \citealp{2010AJ....140.1868W}), which is a joint project of the University of California, Los Angeles, and the Jet Propulsion Laboratory (JPL)/Caltech, funded by NASA. This project was developed in part at the 2017 Heidelberg \emph{Gaia} Sprint, hosted by the Max-Planck-Institut f\"ur Astronomie, Heidelberg. This work has made use of data from the European Space Agency (ESA) mission {\it Gaia} (\url{http://www.cosmos.esa.int/gaia}), processed by the {\it Gaia} Data Processing and Analysis Consortium (DPAC, \url{http://www.cosmos.esa.int/web/gaia/dpac/consortium}). Funding for the DPAC has been provided by national institutions, in particular the institutions participating in the {\it Gaia} Multilateral Agreement. We note that 46 of the \news\ new candidate members presented in this paper were also independently uncovered in Faherty et al. (submitted to ApJ).

\emph{JG} wrote the codes, manuscript, generated figures and led all analyses; \emph{ORL} performed parts of the literature cross-matches and interpretation, and helped generate lists of new candidate members; \emph{JKF} helped parsing young association literature data and provided general comments; \emph{RD} shared comments and supervized \emph{ORL}; and \emph{LM} helped with the construction of color-magnitude sequences.

\software{BANYAN~$\Sigma$ \citep{2018ApJ...856...23G}.}

\bibliographystyle{apj}

\end{document}